\documentclass[pdflatex,sn-nature,iicol]{sn-jnl}

\usepackage{graphicx}
\usepackage{dcolumn}
\usepackage{bm}
\usepackage{amsmath}
\usepackage{amssymb}
\usepackage{braket}
\usepackage[switch]{lineno}

\unnumbered

\begin{document}


\title[Article Title]{Microwave-Induced Optomagnetism in High-Temperature Superconductors}

\author*[1]{\fnm{Anoop} \sur{Dhillon}}\email{a33dhillon@uwaterloo.ca}
\author{\fnm{Amir} \sur{Borji}} \email{aborji@uwaterloo.ca}
\author*[1,2,3]{\fnm{Hamed} \sur{Majedi}} \email{ahmajedi@uwaterloo.ca}

\affil[1]{\orgdiv{Department of Electrical and Computer Engineering}, \orgname{University of Waterloo}, \orgaddress{\street{200 University Avenue West}, \city{Waterloo}, \postcode{N2L 3G1}, \state{Ontario}, \country{Canada}}}
\affil[2]{\orgdiv{Department of Physics and Astronomy}, \orgname{University of Waterloo}, \orgaddress{\street{200 University Avenue West}, \city{Waterloo}, \postcode{N2L 3G1}, \state{Ontario}, \country{Canada}}}
\affil[3]{\orgdiv{Waterloo Institute of Nanotechnology}, \orgaddress{\street{200 University Avenue West}, \city{Waterloo}, \postcode{N2L 3G1}, \state{Ontario}, \country{Canada}}}

\abstract{
    We report the first experimental observation of a steady-state, microwave-driven inverse Faraday effect in a high-temperature superconductor. Circularly polarized microwave radiation generates a helicity-dependent response in an epitaxial $\mathrm{YBa_2Cu_3O_{7-\delta}}$ film, detected using homodyne Hall transport. The optomagnetic response emerges exclusively below $T_c$, vanishes in the normal state, and exhibits no power-dependent counterpart under linearly polarized excitation. The effective optomagnetic conversion reaches $1.75\,\mathrm{T}/(\mathrm{W\,cm^{-2}})$, surpassing optical benchmarks by several orders of magnitude. At higher microwave powers, the signal collapses when the self-generated field exceeds $B_{c1}$, marking the onset of a vortex phase-slip regime, and subsequently re-emerges at mode-locked vortex-washboard harmonics. These results establish steady-state microwave optomagnetism as a route to contactless, non-inductive magnetic control and nonequilibrium vortex spectroscopy in superconducting quantum systems.
}


\maketitle

The inverse Faraday effect (IFE), in which circularly polarized electromagnetic (EM) radiation induces static magnetization \cite{L.P.PitaevskiiElectricForcesTransparent1961}, promises a contactless mechanism for locally generating tunable yet stable magnetic fields without Joule heating or electrical noise. Although the IFE has been demonstrated across a diverse range of materials \cite{V.P.MOpticallyInducedMagnetizationResulting1965, DeschampsInverseFaradayEffect1970, Mangin2014-ad, PhysRevApplied.12.024019, 10.1063/5.0278337}, the induced magnetization has remained weak and transient, confined largely to ultrafast timescales \cite{Ghamsari2016-ey}.

At optical frequencies, the IFE is commonly understood to arise from stimulated population imbalances between electronic states of opposite angular momentum \cite{P.V.MTheoreticalDiscussionInverse1966, K.K.U+UltrafastNonthermalControl2005b}. In the long-wavelength limit, however, the driving field couples to the distributed electronic response of the material \cite{B.B.OQuantumTheoryInverse2014a}, and this nonlinear optical process \cite{PhysRevB.102.214401} manifests macroscopically as the magnetic moment of a radiation-induced circulating current density \cite{pomeau1967effet, Karpman1982215, HertelTheoryInverseFaraday2006}. While this steady-state current is severely suppressed by rapid carrier scattering in normal metals, superconductors support long-lived, dissipationless currents. Superconducting platforms are therefore predicted to host an extraordinarily enhanced low-frequency IFE \cite{MajediMicrowaveInducedInverseFaraday2021}, yet direct experimental observation of this steady-state regime has remained elusive.

Here, we report the first observation of a robust steady-state, low-frequency inverse Faraday effect in the high-temperature superconductor YBa$_2$Cu$_3$O$_{7-\delta}$ (YBCO) driven by $5$ GHz microwave excitation. Although long predicted, detecting its transport signature remains notoriously difficult due to the parasitic rectification and bolometric heating associated with a strong, free-space microwave drive. To overcome these challenges, we designed a novel, low-noise homodyne transport apparatus to isolate the genuine helicity-dependent response. This response emerges abruptly below the superconducting transition temperature ($T_c$) and vanishes in the normal state due to the onset of dissipative transport. Furthermore, we resolve a distinct suppression in the induced Hall signal above the lower critical field ($B_{c1}$), consistent with the proliferation of radiation-induced magnetic vortices, followed by a signal revival when vortex motion becomes mode-locked to the microwave drive. These results establish superconductors as a viable platform for dissipationless magnetic field generation, offering a new paradigm for the non-invasive control of integrated quantum devices, while opening a novel frontier in the spectroscopic investigation of topological phases in quantum materials.

\subsection{Theoretical Framework: Two-Fluid Model and Optomagnetic Response}
Theoretical descriptions of the IFE in superconductors generally fall into two categories: mesoscopic and microscopic. The mesoscopic approach \cite{M.M.T+InverseFaradayEffect2021} extends the time-dependent Ginzburg-Landau theory to the fluctuation regime \cite{HertelTheoryInverseFaraday2006}, tracking the radiation-induced rotation of the condensate current density and its resultant magnetization. Conversely, the microscopic approach \cite{MajediMicrowaveInducedInverseFaraday2021} employs a two-fluid representation to derive the magnetic moments of the coexisting normal and superconducting electron gases under an incident field. This microscopic picture is particularly well-suited for describing the steady-state magnetization induced by sub-gap excitation frequencies ($\hbar\omega \ll 2\Delta$) \cite{osti_4097498}. Although recent theoretical extensions have explored non-equilibrium magnetization via above-gap excitations and localized vortex generation \cite{PhysRevB.110.094302, PhysRevB.110.054506, PLASTOVETS2023129001, PhysRevB.106.174504, 10.1063/5.0165874}, fully characterizing the optomagnetic response of a Type-II superconductor requires synthesizing both frameworks. While the microscopic model cleanly captures the low-field Meissner state, mesoscopic vortex electrodynamics dictate the response once the self-generated field grows.

\begin{figure*}
    \centering
    \includegraphics[width=0.99\textwidth]{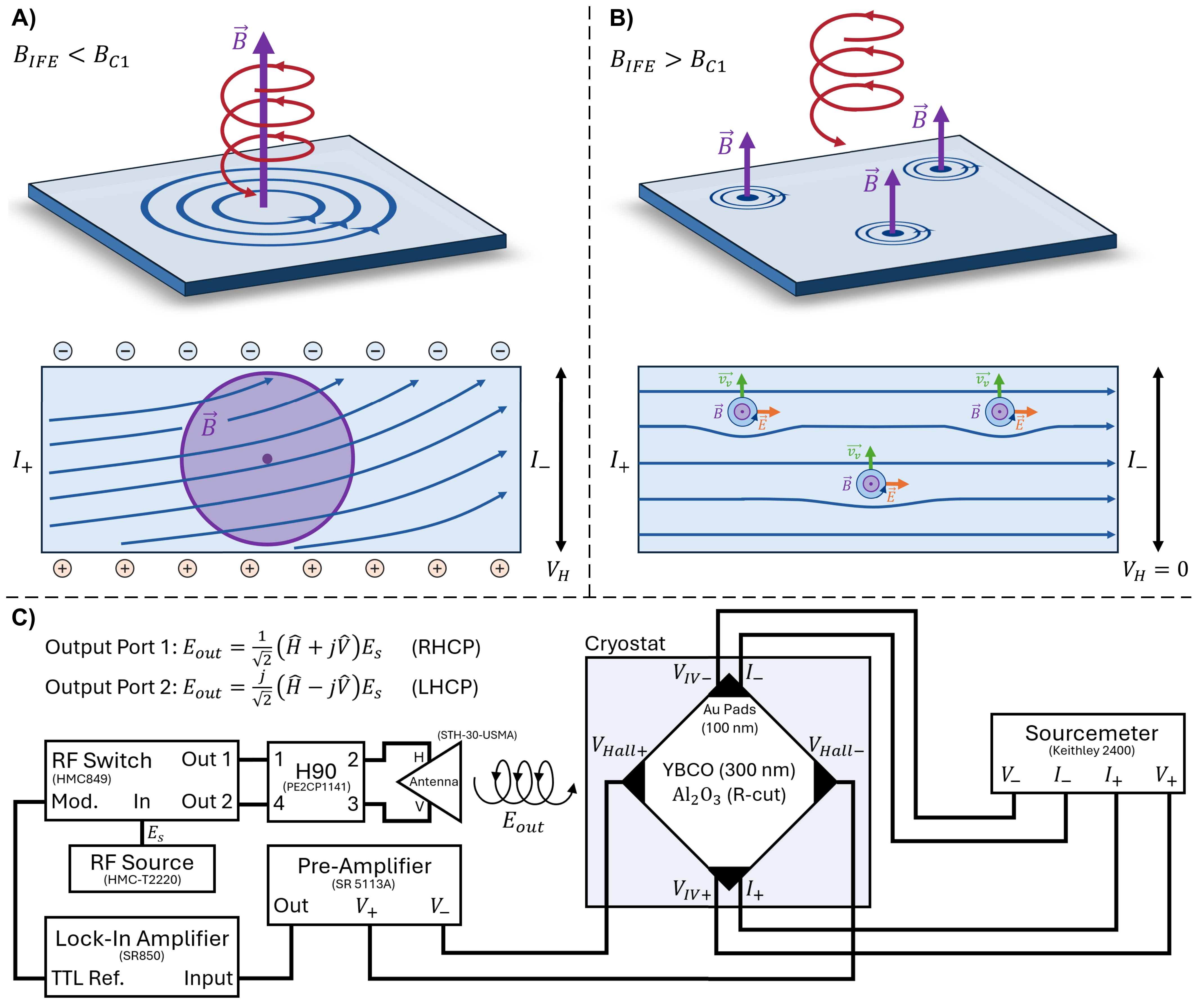}
    \caption{\textbf{Schematics illustrating the optomagnetic response being measured, and the associated experimental apparatus.} \textbf{A)} Optomagnetic response mechanism within the low-field state, showing a distributed global magnetization caused by the magnetic moment of induced circulating current densities, and the associated Hall voltage ($V_H$) formed through the deflection of a bias current under it. \textbf{B)} Optomagnetic response mechanism above the vortex entry threshold field ($B_{c1}$), in which flux is now localized to individual vortices, suppressing the deflection of charge carriers and instead producing a longitudinal phase-slip voltage associated with the deflection of vortices by the induced quantum pressure. \textbf{C)} Experimental apparatus implemented to perform an ultra-low-noise measurement of the Hall voltage. }
    \label{fig:aboveVBelowandApparatus}
\end{figure*}

In the low-field Meissner regime, the superconductor is well-approximated as a two-fluid mixture of superconducting and normal electrons with densities $n_s$ and $n_n$, respectively \cite{osti_4097498}. A circularly polarized electromagnetic wave drives these carriers into local, circular trajectories (Fig. \ref{fig:aboveVBelowandApparatus}A) \cite{MajediMicrowaveInducedInverseFaraday2021}, with an orbital radius defined by the ratio between their angular velocity and the driving frequency $\omega$. Each orbit contributes a local magnetic moment $\mathbf{L}(t)$, yielding a time-dependent magnetization $\mathbf{M}(t)$. Although the net drift velocity of the carriers remains zero, a finite steady-state magnetization $\langle\tilde{\mathbf{M}}\rangle$ emerges upon time-averaging due to the transfer of angular momentum from the incident photons to the orbital motion of the electron fluids. Under coherent driving, the magnitude of this net magnetization is given by \cite{supplement}:
\begin{align}
    \langle\tilde{\mathbf{M}}\rangle=\gamma(\omega)\operatorname{Im}\left[\tilde{\mathbf{E}}\times \tilde{\mathbf{E}}^*\right] \label{eqn:magnetizationIFE}
\end{align}
where the gyration coefficient $\gamma(\omega)$ incorporates the transport characteristics of each electron gas (effective mass $m^*$, scattering time $\tau$, and charge $e$):
\begin{align}
    \gamma(\omega)=\frac{n_se^3}{4{m_s^*}^2 \omega^3}+\frac{n_n e^3}{4{m_n^*}^2\omega^3}\frac{\omega^2 \tau_n^2}{1+\omega^2\tau_n^2} \label{eqn:gyrationCoefficient}
\end{align}
These two terms represent the contribution from the superconducting condensate and normal quasiparticles, respectively. At microwave frequencies, the normal state contribution is suppressed by a factor of $\omega^2 \tau_n^2/(1+\omega^2\tau_n^2)\approx 10^{-7}$ relative to that of the superconducting condensate. This extreme suppression explains why steady-state optomagnetism remains unobservable in normal metals, while highlighting the massive enhancement realized in the superconducting state.

\subsection{Transport Signature and Vortex Dynamics}
To detect this internal magnetization without introducing off-chip magnetic sensors that alter local field profiles, we measure the induced Hall voltage generated across the superconductor under an applied DC bias current. The resulting Hall resistivity $\tilde{\rho}_H$ is modelled via the two-fluid gyromagnetic transport tensor \cite{supplement}:
\begin{align}
    \tilde{\rho}_H=\frac{\tilde{\sigma}_{xy}}{\tilde{\sigma}_{xx}^2+\tilde{\sigma}_{xy}^2}\label{eqn:hallCoeff}
\end{align}
where $\tilde{\sigma}_{xy}=\tilde{\sigma}_{n0}\tilde{\omega}_{cn}\tau_n + \tilde{\sigma}_{s0}\tilde{\omega}_{cs}$ and $\tilde{\sigma}_{xx}=\tilde{\sigma}_{n0} + j\omega_H\tilde{\sigma}_{s0}$. Here, $\tilde{\sigma}_{n0} = (n_n e^2 \tau_n/m_n^*)/(1 + \tilde{\omega}_{cn}^2 \tau_n^2)$ and $\tilde{\sigma}_{s0} = (n_s e^2/m_s^*)/(\tilde{\omega}_{cs}^2-\omega_H^2)$ represent the baseline gyromagnetic conductivities of each electron fluid, characterized by cyclotron frequencies $\tilde{\omega}_{ci}=e\tilde{B}_{\text{IFE}}/m_i^*$ (for $i \in \{n, s\}$) that depend directly on the induced optomagnetic field ($\tilde{B}_{\text{IFE}}=\mu_o \langle\tilde{\mathbf{M}}\rangle$).

To isolate this helicity-dependent response from rectified or bolometric artifacts, the polarization of the incident field is modulated between left- (LHCP) and right-handed circular polarization (RHCP) at a homodyne frequency $\omega_H$, and demodulated via lock-in detection (Fig. \ref{fig:aboveVBelowandApparatus}C). Modulating the induced magnetic field at an intermediate frequency enables a non-zero Hall voltage measurement ($\tilde{V}_H = \int \tilde{\boldsymbol{\rho}}_t \mathbf{J} \cdot d\mathbf{l}$), even when driven by a DC bias current.

When the induced optomagnetic field exceeds the lower critical field threshold $B_{c1}$, the system transitions from the uniform Meissner state to a mixed state where vortex generation becomes energetically favourable \cite{Abrikosov:1956sx}. The spatially homogeneous magnetization of the electronic orbits is then replaced by discrete, localized flux penetration within vortex cores (Fig. \ref{fig:aboveVBelowandApparatus}B) \cite{M.M.T+InverseFaradayEffect2021}.

In this vortex regime, the equilibrium two-fluid hydrodynamics breaks down \cite{osti_4097498}, and normal quasiparticles outside the vortex cores no longer experience a uniform internal magnetic field, causing the spatial profile of the optomagnetic supercurrents to be fundamentally disrupted. Consequently, the conventional transverse Hall voltage predicted by equation (\ref{eqn:hallCoeff}) is suppressed. Although the vortices experience an orthogonal Magnus force relative to the bias current, arising from quantum pressure differentials \cite{meanFieldVortices_chapman}, as detailed in the Supplementary Information \cite{supplement}, they carry no net electric charge, meaning their transverse deflection does not generate a standard transverse electrostatic Hall potential. Instead, the motion of these localized magnetic flux tubes generates a longitudinal phase-slip electric field parallel to the bias current \cite{meanFieldVortices_chapman}. Therefore, once the radiation-induced optomagnetic field crosses the threshold for vortex entry ($B_{c1}$), the measured Hall voltage decays rapidly toward zero, even as the total local magnetization of the film remains large \cite{guidedVortexMotion_Wordenweber}.

\subsection{Observation of the Superconducting Inverse Faraday Effect}
\begin{figure*}
    \centering
    \includegraphics[width=0.99\textwidth]{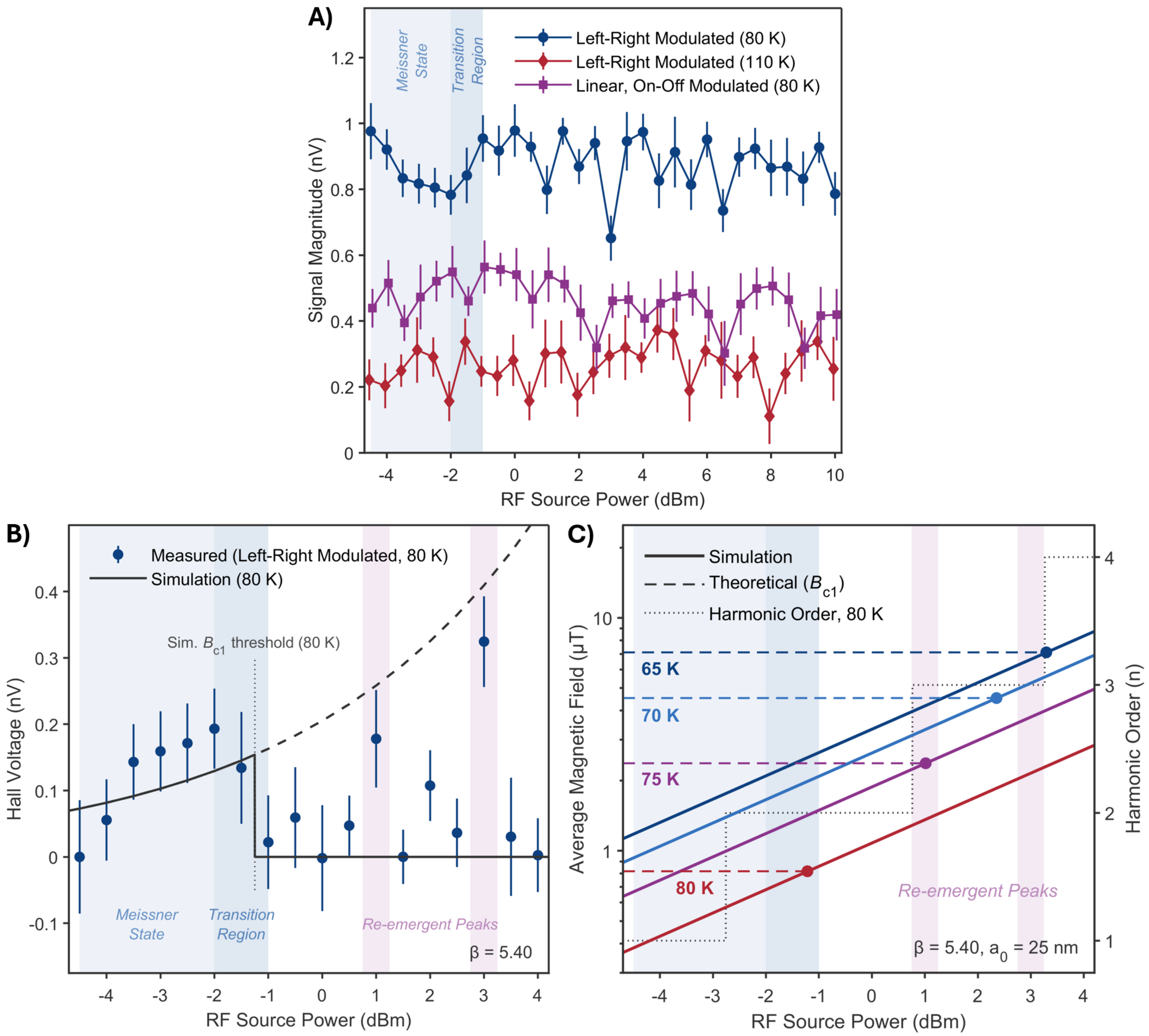}
    \caption{\textbf{Experimental observation of the superconducting optomagnetic response and vortex crossover.} Shaded vertical regions track the distinct physical regimes across power sweeps below $T_c$: the linear Meissner state (left blue band) and the vortex entry transition zone (right blue band). \textbf{A)} Measured lock-in signal magnitude capturing the helicity-dependent Hall voltage difference under left- versus right-handed circularly polarized microwave excitation (Left-Right Modulated) recorded below $T_c$ at $80$ K (blue circles) and above $T_c$ at $110$ K (red diamonds), alongside the linear polarization control at $80$ K (Linear, On-Off Modulated, purple squares). \textbf{B)} Zoomed-in helicity-dependent experimental data below $T_c$ (blue circles, referenced to the $-4.5$ dBm baseline) compared directly against numerical simulations (Ansys HFSS and COMSOL, solid black line). The modified simulation incorporates manual zeroing of the Hall signal above the lower critical field ($B_{c1}$) threshold (vertical dotted line) to account for vortex entry truncation, and utilizes an empirical local electric field enhancement factor ($|\tilde{\mathbf{E}}_0|=\beta|\tilde{\mathbf{E}}_{0i}|$, where $\beta=5.40$) to match low-power experimental signal magnitudes. The unattenuated continuation of the simulation is tracked via a dashed black line, highlighting its close alignment with the Re-emergent Peaks (pink shading) observed at higher source powers. \textbf{C)} Simulated average internal magnetic field strength (left axis) plotted against the theoretical vortex entry field threshold ($B_{c1}$) (horizontal dashed lines) below $T_c$. The same local electric field enhancement factor ($\beta=5.40$) aligns the simulated field crossover at $T=80$ K (intersection marker) with the experimental suppression threshold in (\textbf{B}). Overlaid is the calculated vortex washboard harmonic order $n=f_{\text{wb}}/f_{\text{mw}}$, where $n\in\mathbb{Z}$ (right axis, dotted line), evaluated using the nominal pinning length of $a_0=25$ nm, determined through vortex pinning saturation measurements. The 3rd and 4th harmonics of the washboard frequency are seen to correspond with the re-emergence of the transport signature, confirming the presence of a dynamic, coherent, phase-locked vortex regime. Data points in (\textbf{A} and \textbf{B}) correspond to the means of block-averaged time-series data taken across statistically independent segmentation periods; error bars represent the standard error of the mean (SEM).}    \label{fig:simRealComparison}
\end{figure*}

To detect the optomagnetic response, we monitor the helicity-dependent Hall voltage $V_H$ (equation (\ref{eqn:hallCoeff})) under intense microwave excitation while applying a continuous DC bias across the YBCO sample (Fig. \ref{fig:aboveVBelowandApparatus}C). At temperatures well below the superconducting transition ($T=80$ K $<T_c$), sweeping the incident microwave power from $-4.5$ to $-2$ dBm reveals a robust, monotonic change in the measured lock-in signal magnitude, establishing a clear trend well outside the standard error of the mean (SEM) of the data (Fig. \ref{fig:simRealComparison}A). This signal scales in strict accordance with the steady-state magnetization predicted by equation (\ref{eqn:magnetizationIFE}). Crucially, this power-dependent response is entirely absent in control measurements performed with linearly polarized radiation (On-Off Modulated), ruling out parasitic contributions from microwave rectification or bolometric heating.

To isolate the intrinsic optomagnetic response from baseline instrumentation noise, we reference the power-dependent signal against the $-4.5$ dBm baseline data point and compare the resulting scaling directly to numerical transport simulations (Ansys HFSS and COMSOL). While the experimental and simulated trends exhibit excellent qualitative agreement (Fig. \ref{fig:simRealComparison}B), the measured Hall voltage exceeds bare two-fluid calculations by a factor of $29.2$. This corresponds to a local electric field enhancement factor of $\beta=5.40$ (where, $|\tilde{\mathbf{E}}_0|=\beta|\tilde{\mathbf{E}}_{0i}|$), as the transverse voltage scales quadratically with field amplitude ($V\propto B\propto\beta^2|\tilde{\mathbf{E}}_{0i}|\approx29.2 |\tilde{\mathbf{E}}_{0i}|$). This enhancement is primarily attributed to strong geometric field concentration effects inherent to high-aspect-ratio thin films ($d=300$ nm, $w=18$ mm) under a long-wavelength microwave drive ($\lambda=6$ cm) \cite{PhysRevB.78.054407}, supplemented by localized field buildup near the corner gold contact pads and wirebonds. Detailed simulation parameters are provided in the Supplementary Information \cite{supplement}.

The statistical validity of this Meissner-state response is exceptionally robust. As highlighted in Fig. \ref{fig:simRealComparison}B, the baseline-referenced signals from $-3.5$ dBm to $-2$ dBm possess lower standard error of the mean (SEM) boundaries that remain clearly separated from the upper statistical bound of the $-4.5$ dBm baseline. This distinct separation confirms that the low-power signal represents a genuine field-induced transport phenomenon rather than stochastic noise.

The superconducting origin of this response is further confirmed by its thermal evolution. Upon warming the sample well above $T_c$, to $110$ K, where normal-state quasiparticle scattering dominates, the helicity-dependent signal collapses completely into the system noise floor, with error bars continuously overlapping the baseline (Fig. \ref{fig:simRealComparison}A). This extinction agrees quantitatively with equation (\ref{eqn:gyrationCoefficient}), which dictates a $10^{-7}$ suppression of the gyration coefficient due to rapid carrier scattering once the dissipationless condensate is destroyed.

\subsection{Vortex Breakdown and High-Power Suppression}
When the incident microwave power is increased beyond $-2$ dBm, the measured Hall voltage deviates from its monotonic growth, decaying rapidly toward the baseline and vanishing entirely by $-1$ dBm (Fig. \ref{fig:simRealComparison}A). Above $-1$ dBm, the data points lose their coherent power dependence and cluster around zero within statistical error, indicating a return to a noise-dominated regime (Fig. \ref{fig:simRealComparison}B). This abrupt collapse signifies that the self-generated optomagnetic field has exceeded the lower critical field threshold $B_{c1}$. Above $B_{c1}$, the entry of quantized magnetic vortices breaks the spatial uniformity of the Meissner state, localizing field penetration through their cores. Through their motion, localized phase-slip dynamics are introduced, which quench the macroscopically averaged Hall voltage toward zero \cite{guidedVortexMotion_Wordenweber}, even as the magnetization of the sample remains high.

This understanding is independently supported by comparing the simulated internal magnetic field generated within the YBCO film against the material's vortex entry field (Fig. \ref{fig:simRealComparison}C). Using the same electric field enhancement factor ($\beta=5.40$) derived from low-power transport matching, the calculated optomagnetic field crosses the $B_{c1}$ threshold precisely within the $-2$ to $-1$ dBm transition window at the point where the optomagnetic signature is first suppressed with statistical significance. This quantitative alignment confirms that the optomagnetically generated field is sufficiently intense to self-nucleate a vortex lattice. At the device level, comparing this $B_{c1}$ threshold at $80$ K ($0.82$ $\mu$T; \ref{fig:simRealComparison}C) directly to the incident free-space microwave intensity at which it occurs ($0.468$ $\mu\text{W}/\text{cm}^2$ at an RF source power of $-1.25$ dBm), yields an exceptionally large effective optomagnetic response of $\gamma_{\text{eff}}(\omega)=1.75$ $\text{T}/(\text{W}/\text{cm}^2)$ (or $49.5$ $\mu\text{T}/\text{W}$). This conversion efficiency exceeds optical-frequency IFE benchmarks, which are only observable under ultrafast, high-energy excitation \cite{10.1021/acsnano.1c06922}, by several orders of magnitude, driven by the absence of dissipative scattering in the superconducting state. Factoring out the local geometric intensity enhancement ($\beta^2=29.2$) isolates an intrinsic material gyration coefficient of $\gamma(\omega)=59.9$ $\text{mT}/(\text{W}/\text{cm}^2)$ (or $1.7$ $\mu\text{T}/\text{W}$), which quantifies the fundamental light-matter coupling efficiency of the superconducting fluid.

Notably, this suppressed baseline is interrupted by two statistically distinct, re-emergent peaks at $+1$ and $+3$ dBm that rise well above noise and track the unattenuated continuation of the Meissner transport curve (Fig. \ref{fig:simRealComparison}B). As shown in Fig. \ref{fig:simRealComparison}C, these peaks are not random fluctuations; rather, they align precisely with microwave powers where the vortex lattice washboard frequency matches higher-order integer harmonics ($n=3,4$) of the microwave drive ($f_{\text{wb}}=nf_{\text{mw}}$). At higher powers ($>+4$ dBm), local thermal loading begins to set in, as evidenced by the downward trend in the linear polarization control data (Fig. \ref{fig:simRealComparison}A). This thermal dissipation degrades the local condensate density, disrupting the global phase coherence required for complete mode-locking. Consequently, the $n=5$ harmonic ($+5.2$ dBm) is entirely suppressed, while the $n=6$ harmonic at $+6.5$ dBm exhibits only a weak, attenuated re-emergence. Crucially, the $n=1$ and $n=2$ washboard harmonics fall strictly below the $B_{c1}$ field threshold, explaining their absence: in that regime, vortices have not yet entered the film, so no washboard resonance can occur.

\subsection{Discussion and Conclusion}

The generation of a steady-state optomagnetic response in the low-field regime highlights a fundamental distinction between equilibrium diamagnetism and driven non-equilibrium electrodynamics. Unlike the static Meissner state, governed by equilibrium energy minimization, our sub-gap continuous-wave microwave excitation ($\hbar\omega\ll 2\Delta$) establishes a non-equilibrium steady state wherein the intact condensate supports driven phase dynamics. Furthermore, because the film thickness ($d=300$ nm) is much smaller than the in-plane penetration depth ($\lambda_{ab,80\text{ K}}=693.7$ nm), magnetic screening is governed by Pearl electrodynamics \cite{PearlCURRENTDISTRIBUTIONSUPERCONDUCTING1964} rather than bulk London screening. Consequently, supercurrents and magnetic fields permeate the entire film volume.

The dynamic evolution of this state above the lower critical field ($B_{c1}$) offers direct insight into the non-equilibrium electrodynamics of driven flux lattices \cite{PhysRevB.111.214510}. Immediately above $B_{c1}$, the uniform magnetic field condenses into localized Abrikosov-Pearl flux tubes \cite{PearlCURRENTDISTRIBUTIONSUPERCONDUCTING1964}, quenching the macroscopic Hall voltage toward zero \cite{guidedVortexMotion_Wordenweber}. Microscopically, this suppression arises because quasiparticle transport becomes strongly scattered by the steep phase gradients ($\nabla\phi$) surrounding individual vortex cores \cite{guidedVortexMotion_Wordenweber}. Under unsynchronized driving from the microwave field and DC bias current, vortices undergo disordered, turbulent motion across the microscopic pinning landscape. The resulting asynchronous $2\pi$ phase-slip events generate chaotic transverse potentials that average out to zero across the sample ($\langle V_{H}\rangle=0$).

However, as the microwave driving power is increased, the vortex drift velocity ($v_v$) scales accordingly. At specific power thresholds, $v_v$ satisfies a resonance condition where vortices travel an integer number ($n$) of pinning lattice spacings ($a_0$) per microwave period ($T_{\text{mw}}$), such that $v_v=na_0/T_{\text{mw}}$ \cite{PhysRevLett.74.3684}. This condition locks the washboard transit frequency ($f_{\text{wb}}=v_v/a_0$) to an integer harmonic of the driving field ($f_{\text{wb}}=nf_{\text{mw}}$). Using the characteristic pinning spacing $a_0=25$ nm, determined from the matching field ($B_\Phi$) plateau in critical current measurements ($J_c(B)$) \cite{CriticalCurrentCharacterisation2023}, and the local field enhancement factor $\beta=5.40$, we find that the $n=3$ and $n=4$ washboard harmonics align precisely with the re-emergent Hall peaks at $+1$ dBm and $+3$ dBm (Fig. \ref{fig:simRealComparison}C). At these resonant thresholds, the vortex lattice mode-locks into a highly ordered, synchronously sliding state \cite{Mallayya2024}. This spatiotemporal coherence transforms chaotic phase-slip noise into a periodic voltage waveform with a net steady-state component, restoring the unattenuated Hall voltage. This sharp collapse and structured harmonic recovery constitute a definitive fingerprint of Type-II vortex dynamics.

This framework yields a concrete, testable prediction. Because $B_{c1}$ increases at lower temperatures, the threshold power required for vortex nucleation must shift upward (from $-1.25$ dBm at $80$ K to a predicted $+3.25$ at $65$ K; Fig. \ref{fig:simRealComparison}C). Replicating these power sweeps across lower temperature ranges will therefore predictably shift the onset of Hall voltage suppression to higher drive powers. Furthermore, because $v_v$ is governed by the temperature-dependent flux-flow viscosity $\eta(T)$ and microwave-induced current density $J_{\text{mw}}(T)$ \cite{supplement}, the power thresholds matching washboard harmonics will display distinct thermal scaling for samples in the thin-film Pearl regime. Verifying these scaling laws via temperature-resolved transport or direct magnetic imaging, such as scanning NV-centre \cite{PhysRevApplied.10.034032} or SQUID magnetometry \cite{PhysRevLett.92.157006}, will map the complete non-equilibrium optomagnetic phase diagram of Type-II superconductors.

Beyond its fundamental implications for light-matter interactions in strongly correlated materials, this low-frequency, dissipationless optomagnetic effect addresses a critical engineering bottleneck in quantum technology architectures. Standard superconducting quantum circuits rely on inductive control lines for magnetic biasing, which introduce thermal loads, cross-talk, and high-frequency decoherence \cite{GennesSuperconductivityMetalsAlloys2018, Schlosshauer-SelbachDecoherenceQuantumtoclassicalTransition2007}. By demonstrating that microwave fields can dynamically induce stable, localized internal magnetic fields without resistive losses, our findings establish a scalable pathway toward non-thermal coherent control of superconducting qubits and topological devices.

\section{Methods}
\subsection{Sample Preparation and Device Architecture}
\begin{figure*}
    \centering
    \includegraphics[width=0.5\textwidth]{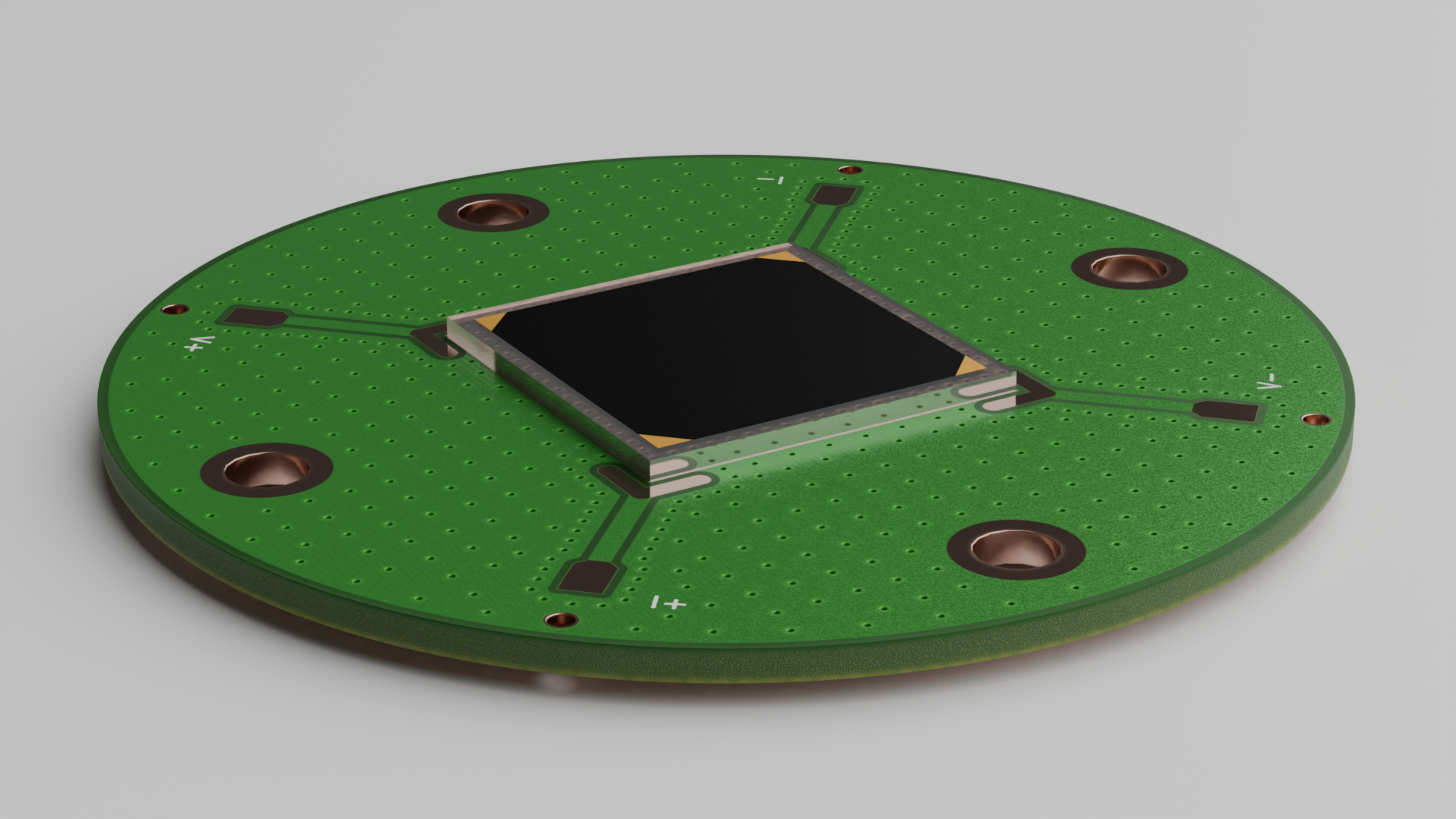}
    \caption{\textbf{Three-dimensional rendering of the device and electrical transport layout.} Perspective view of the epitaxial YBCO thin film on its sapphire substrate, wire-bonded to the host PCB for transverse Hall voltage measurements.}
    \label{fig:sampleIllustration}
\end{figure*}

The experiments were performed on a commercial, high-quality M-Type $\mathrm{YBa_2Cu_3O_{7-\delta}}$ (YBCO) thin film (thickness $d=300$ nm) epitaxially grown in an $18\times18$ mm$^2$ pattern on an r-cut sapphire substrate (Ceraco). The M-Type YBCO film used here is deliberately engineered with a random distribution of Y$_2$O$_3$ nanoparticles to enhance microwave power handling via chemical pinning. To facilitate electrical transport measurements in a Van der Pauw configuration, triangular gold contact pads (thickness $100$ nm, lateral dimensions $1.8$ mm) were deposited onto the corners of the film. An electrical connection between the sample and an electroless nickel immersion gold (ENIG)-coated printed circuit board (PCB) was established via aluminum wedge-bonding (Fig. \ref{fig:sampleIllustration}). In this configuration, a $1$ mA current is injected along one diagonal using a sourcemeter (Keithley 2400), while the Hall voltage is read along the other. A four-probe measurement of the sample's resistance is also performed along the current-carrying diagonal, to identify the onset of the superconducting state.

\subsection{Cryogenic Microwave Setup and Homodyne Modulation}
The sample-mounted PCB was housed within an evacuated, liquid-nitrogen-cooled cryostat. Undistorted high-frequency EM coupling was achieved via a $20$ cm diameter microwave-transparent acrylic aperture, permitting low-loss transmission at $5$ GHz from an external dual-linearly polarized horn antenna (RF Elements, STH-30-USMA, gain at boresight: $18.27$ dBi) positioned at a distance of $d=0.53$ m.

The helicity of this radiation was modulated using an RF switch (Microwave system path loss: $4.84$ dB), controlling which port of a Hybrid 90$^\circ$ coupler was fed, resulting in either a $+90^\circ$ or $-90^\circ$ phase shift between the quadratures of the antenna, corresponding to a right- or left-handed circularly polarized output, as described in the inset text of Fig. \ref{fig:aboveVBelowandApparatus}C. Modulating this helicity at the homodyne frequency ($f_{H}=83.099$ kHz) set by a lock-in amplifier (Stanford Research Systems, SR850) allowed only those components of the measured Hall voltage signal that oscillate in response to that specific excitation to be extracted, removing artifacts related to rectification or bolometric heating. Notably, while the microsecond period of this modulation is dynamic, it operates several orders of magnitude slower than the picosecond intrinsic relaxation timescales of the superconducting and normal electrons. The configuration therefore probes a quasi-steady state wherein the electronic system achieves local thermodynamic equilibrium during each polarization cycle, effectively isolating steady-state optomagnetic behaviour from transient artifacts. To further rule out power-dependent effects, a control measurement was conducted in which the Hybrid 90$^\circ$ coupler was bypassed, and the RF switch was used to modulate the antenna between a driven and non-driven state via the horizontally linearly polarized ($\hat{H}$) port, with the other port matched.

\subsection{Signal Conditioning and Noise Mitigation}
To extract the minuscule, helicity-dependent optomagnetic signature while suppressing common-mode electromagnetic interference and ambient microwave pickup, the induced Hall voltage was measured differentially across the Van der Pauw geometry. The raw differential signal was routed through a low-noise differential voltage preamplifier (Signal Recovery 5113A, input noise voltage: $4$ nV/$\sqrt{\text{Hz}}$, Gain: $10,000$) before demodulation at the lock-in amplifier. Ground loops and instrumentation noise were minimized by star-grounding the cryostat chassis and all peripheral AC equipment (excluding the floating preamplifier) to a central earth terminal using high-conductivity copper braiding.

\subsection{Data Acquisition and Statistical Analysis}
Lock-in detection was performed using a second-order low-pass filter (effective noise bandwidth: $42$ mHz) with a time constant of $\tau=3$ s. For each experimental power and temperature data point presented in Fig. \ref{fig:simRealComparison}, a continuous time series was recorded over a total duration of $100\tau$. To ensure the system has reached a steady state, the initial $10\tau$ segment of each time series was discarded. The remaining data were partitioned and block-averaged in segments exceeding the filter's characteristic correlation time ($10\tau$) to guarantee statistically independent samples. The final data points and their associated error bars (Fig. \ref{fig:simRealComparison}A) denote the calculated mean and standard error of the mean (SEM) across these independent blocks.

\backmatter

\bmhead{Funding Statements}
The authors acknowledge the support of the Natural Sciences and Engineering Research Council of Canada (NSERC), the Discovery Grant program, RGPIN-2024-04522.

\bmhead{Supplementary information}
A comprehensive justification of the theoretical developments presented in this paper can be found in the Supplementary Information, along with additional computational simulations and supplementary experimental results.

\noindent


\end{document}